\RequirePackage{lineno}
\documentclass[aps,twocolumn,superscriptaddress,showpacs,preprintnumbers,amsmath,amssymb]{revtex4}
\usepackage{graphicx,color,dcolumn,booktabs,bm}
\usepackage{longtable,lscape}
\usepackage{amsmath}
\usepackage{indentfirst}
\usepackage{epsfig}
\usepackage{feynmf}   
\usepackage{epstopdf}   
\usepackage{slashed}  
\usepackage{cases}
\usepackage{color}
\usepackage{multirow}
\usepackage{graphicx,color,dcolumn,booktabs,bm}
\usepackage{verbatim}
\usepackage[colorlinks,linkcolor=red,anchorcolor=green,citecolor=blue]{hyperref}
\usepackage{mathrsfs}
\usepackage{rotating}
\usepackage{threeparttable}
\usepackage{lineno}

\newcommand{\yfos}{\Upsilon(4S)}
\newcommand{\yfis}{\Upsilon(5S)}

\newcommand{\BR}{{\cal B}}

\newcommand{\ks}{K_S^0}

\newcommand{\EE}{e^+e^-}

\newcommand{\bbc}{B^+ B^-}
\newcommand{\bbn}{B^0\bar{B}^0}

\newcommand{\infb}{\rm fb^{-1}}

\newcommand{\gev}{\rm GeV}

\newcommand{\beq}{\begin{equation}}
\newcommand{\eeq}{\end{equation}}
\newcommand{\bitm}{\begin{itemize}}
\newcommand{\eitm}{\end{itemize}}

\begin{document}


\title{\quad\\[0.1cm] Observation of a vector charmoniumlike state in \boldmath{$e^+e^- \to D^+_sD_{s1}(2536)^-+c.c.$}}

\noaffiliation
\affiliation{University of the Basque Country UPV/EHU, 48080 Bilbao}
\affiliation{Beihang University, Beijing 100191}
\affiliation{Brookhaven National Laboratory, Upton, New York 11973}
\affiliation{Budker Institute of Nuclear Physics SB RAS, Novosibirsk 630090}
\affiliation{Faculty of Mathematics and Physics, Charles University, 121 16 Prague}
\affiliation{Chonnam National University, Gwangju 61186}
\affiliation{University of Cincinnati, Cincinnati, Ohio 45221}
\affiliation{Deutsches Elektronen--Synchrotron, 22607 Hamburg}
\affiliation{Key Laboratory of Nuclear Physics and Ion-beam Application (MOE) and Institute of Modern Physics, Fudan University, Shanghai 200443}
\affiliation{Justus-Liebig-Universit\"at Gie\ss{}en, 35392 Gie\ss{}en}
\affiliation{SOKENDAI (The Graduate University for Advanced Studies), Hayama 240-0193}
\affiliation{Gyeongsang National University, Jinju 52828}
\affiliation{Department of Physics and Institute of Natural Sciences, Hanyang University, Seoul 04763}
\affiliation{University of Hawaii, Honolulu, Hawaii 96822}
\affiliation{High Energy Accelerator Research Organization (KEK), Tsukuba 305-0801}
\affiliation{J-PARC Branch, KEK Theory Center, High Energy Accelerator Research Organization (KEK), Tsukuba 305-0801}
\affiliation{Forschungszentrum J\"{u}lich, 52425 J\"{u}lich}
\affiliation{IKERBASQUE, Basque Foundation for Science, 48013 Bilbao}
\affiliation{Indian Institute of Science Education and Research Mohali, SAS Nagar, 140306}
\affiliation{Indian Institute of Technology Guwahati, Assam 781039}
\affiliation{Indian Institute of Technology Hyderabad, Telangana 502285}
\affiliation{Indian Institute of Technology Madras, Chennai 600036}
\affiliation{Indiana University, Bloomington, Indiana 47408}
\affiliation{Institute of High Energy Physics, Chinese Academy of Sciences, Beijing 100049}
\affiliation{Institute of High Energy Physics, Vienna 1050}
\affiliation{Institute for High Energy Physics, Protvino 142281}
\affiliation{INFN - Sezione di Napoli, 80126 Napoli}
\affiliation{INFN - Sezione di Torino, 10125 Torino}
\affiliation{Advanced Science Research Center, Japan Atomic Energy Agency, Naka 319-1195}
\affiliation{J. Stefan Institute, 1000 Ljubljana}
\affiliation{Institut f\"ur Experimentelle Teilchenphysik, Karlsruher Institut f\"ur Technologie, 76131 Karlsruhe}
\affiliation{Kennesaw State University, Kennesaw, Georgia 30144}
\affiliation{King Abdulaziz City for Science and Technology, Riyadh 11442}
\affiliation{Korea Institute of Science and Technology Information, Daejeon 34141}
\affiliation{Korea University, Seoul 02841}
\affiliation{Kyoto University, Kyoto 606-8502}
\affiliation{Kyungpook National University, Daegu 41566}
\affiliation{LAL, Univ. Paris-Sud, CNRS/IN2P3, Universit\'{e} Paris-Saclay, Orsay 91898}
\affiliation{\'Ecole Polytechnique F\'ed\'erale de Lausanne (EPFL), Lausanne 1015}
\affiliation{P.N. Lebedev Physical Institute of the Russian Academy of Sciences, Moscow 119991}
\affiliation{Liaoning Normal University, Dalian 116029}
\affiliation{Faculty of Mathematics and Physics, University of Ljubljana, 1000 Ljubljana}
\affiliation{Ludwig Maximilians University, 80539 Munich}
\affiliation{Luther College, Decorah, Iowa 52101}
\affiliation{Malaviya National Institute of Technology Jaipur, Jaipur 302017}
\affiliation{University of Maribor, 2000 Maribor}
\affiliation{Max-Planck-Institut f\"ur Physik, 80805 M\"unchen}
\affiliation{School of Physics, University of Melbourne, Victoria 3010}
\affiliation{University of Mississippi, University, Mississippi 38677}
\affiliation{University of Miyazaki, Miyazaki 889-2192}
\affiliation{Moscow Physical Engineering Institute, Moscow 115409}
\affiliation{Moscow Institute of Physics and Technology, Moscow Region 141700}
\affiliation{Graduate School of Science, Nagoya University, Nagoya 464-8602}
\affiliation{Kobayashi-Maskawa Institute, Nagoya University, Nagoya 464-8602}
\affiliation{Universit\`{a} di Napoli Federico II, 80055 Napoli}
\affiliation{Nara Women's University, Nara 630-8506}
\affiliation{National Central University, Chung-li 32054}
\affiliation{National United University, Miao Li 36003}
\affiliation{Department of Physics, National Taiwan University, Taipei 10617}
\affiliation{H. Niewodniczanski Institute of Nuclear Physics, Krakow 31-342}
\affiliation{Nippon Dental University, Niigata 951-8580}
\affiliation{Niigata University, Niigata 950-2181}
\affiliation{University of Nova Gorica, 5000 Nova Gorica}
\affiliation{Novosibirsk State University, Novosibirsk 630090}
\affiliation{Osaka City University, Osaka 558-8585}
\affiliation{Pacific Northwest National Laboratory, Richland, Washington 99352}
\affiliation{Panjab University, Chandigarh 160014}
\affiliation{Peking University, Beijing 100871}
\affiliation{University of Pittsburgh, Pittsburgh, Pennsylvania 15260}
\affiliation{Theoretical Research Division, Nishina Center, RIKEN, Saitama 351-0198}
\affiliation{University of Science and Technology of China, Hefei 230026}
\affiliation{Seoul National University, Seoul 08826}
\affiliation{Showa Pharmaceutical University, Tokyo 194-8543}
\affiliation{Soongsil University, Seoul 06978}
\affiliation{University of South Carolina, Columbia, South Carolina 29208}
\affiliation{Sungkyunkwan University, Suwon 16419}
\affiliation{School of Physics, University of Sydney, New South Wales 2006}
\affiliation{Department of Physics, Faculty of Science, University of Tabuk, Tabuk 71451}
\affiliation{Tata Institute of Fundamental Research, Mumbai 400005}
\affiliation{Department of Physics, Technische Universit\"at M\"unchen, 85748 Garching}
\affiliation{Toho University, Funabashi 274-8510}
\affiliation{Earthquake Research Institute, University of Tokyo, Tokyo 113-0032}
\affiliation{Department of Physics, University of Tokyo, Tokyo 113-0033}
\affiliation{Tokyo Institute of Technology, Tokyo 152-8550}
\affiliation{Tokyo Metropolitan University, Tokyo 192-0397}
\affiliation{Virginia Polytechnic Institute and State University, Blacksburg, Virginia 24061}
\affiliation{Wayne State University, Detroit, Michigan 48202}
\affiliation{Yamagata University, Yamagata 990-8560}
\affiliation{Yonsei University, Seoul 03722}

\author{S.~Jia}\affiliation{Beihang University, Beijing 100191} 
\author{C.~P.~Shen}\affiliation{Key Laboratory of Nuclear Physics and Ion-beam Application (MOE) and Institute of Modern Physics, Fudan University, Shanghai 200443} 
\author{C.~Z.~Yuan}\affiliation{Institute of High Energy Physics, Chinese Academy of Sciences, Beijing 100049} 
\author{X.~L.~Wang}\affiliation{Key Laboratory of Nuclear Physics and Ion-beam Application (MOE) and Institute of Modern Physics, Fudan University, Shanghai 200443} 
  \author{I.~Adachi}\affiliation{High Energy Accelerator Research Organization (KEK), Tsukuba 305-0801}\affiliation{SOKENDAI (The Graduate University for Advanced Studies), Hayama 240-0193} 
  \author{H.~Aihara}\affiliation{Department of Physics, University of Tokyo, Tokyo 113-0033} 
  \author{D.~M.~Asner}\affiliation{Brookhaven National Laboratory, Upton, New York 11973} 
  \author{H.~Atmacan}\affiliation{University of South Carolina, Columbia, South Carolina 29208} 
  \author{V.~Aulchenko}\affiliation{Budker Institute of Nuclear Physics SB RAS, Novosibirsk 630090}\affiliation{Novosibirsk State University, Novosibirsk 630090} 
  \author{R.~Ayad}\affiliation{Department of Physics, Faculty of Science, University of Tabuk, Tabuk 71451} 
% ...
\collaboration{The Belle Collaboration}

\begin{abstract}

Using a data sample of 921.9 fb$^{-1}$ collected with the Belle detector, we study the process of $e^+e^-\to D^+_sD_{s1}(2536)^-+c.c.$ via initial-state radiation. We report the first observation of a vector charmoniumlike state decaying to $D^+_sD_{s1}(2536)^-+c.c.$ with a significance of 5.9$\sigma$,  including the systematic uncertainties. The measured mass and width are $(4625.9^{+6.2}_{-6.0}({\rm stat.})\pm0.4({\rm syst.}))~{\rm MeV}/c^{2}$ and $(49.8^{+13.9}_{-11.5}({\rm stat.})\pm4.0({\rm syst.}))~{\rm MeV}$, respectively. The product of the $e^+e^-\to D^+_sD_{s1}(2536)^-+c.c.$ cross section and the branching fraction of $D_{s1}(2536)^-\to{\bar D}^{*0}K^-$ is measured from the $D_s \bar{D}_{s1}(2536)$ threshold to 5.59~GeV.

\end{abstract}


\maketitle

In the past decade, measurements of the exclusive cross sections
for $e^+e^-$ annihilation into charmed or charmed-strange meson
pairs and charmed baryon pairs above the open-charm threshold have\emph{}
attracted much attention~\cite{D77,98,100,D80,101,D83,D76,D79,D82,D072001,weiping}.
Unexpected strong near-threshold enhancements are present in the $e^+e^-\to D \bar D$,
$D^{(*)+}D^{*-}$, $D^+_sD^-_s$, and $\Lambda^+_{c}
\bar{\Lambda}^-_{c}$ cross sections~\cite{D77,98,101,D83,weiping}.
These open-charm final states are dominantly produced from the OZI-allowed strong decays of excited vector charmonium states ($\psi$ states). Due to the lack of adequate experimental
measurements and sophisticated theoretical models, the couplings of these $\psi$ states to the open-charm final states as well as the resonant parameters of these charmonium states are not well measured.

Many additional $Y$ states with $J^{PC} = 1^{--}$ lying with masses above the open-charm threshold have been discovered in the last 14 years~\cite{95,102003,96,99,252002,092001,L98,L99,032004}.
In $e^+e^-\to Y \to \pi^+\pi^-J/\psi$ and $\pi^+\pi^-\psi(2S)$ ($Y=Y(4260)$, $Y(4660)$) processes, events in $\pi^+\pi^-$ mass spectra tend to accumulate at the
$f_{0}(980)$ nominal mass, which has an $s\bar s$ component. Thus, it
is natural to search for $Y$ states with a $(c{\bar s})({\bar c}s)$ quark component. As mentioned in Ref.~\cite{N954}, bound states of $D_s\bar D_s$ mesons, e.g., $D_s\bar
D_{s1}(2536)$, can appear as a result of $f_{0}(980)$ exchange.
Unfortunately, open-charmed-strange production
associated with these $Y$ states has not yet been observed.

In this Letter, we perform an exclusive cross section measurement of $e^+e^-\to
D^+_sD_{s1}(2536)^-~(D_{s1}(2536)^-\to {\bar D}^{*0}K^-/D^{*-}K^0_S)$
as a function of center-of-mass (C.M.) energy from the $D^+_sD_{s1}(2536)^-$ mass threshold
to 5.59~GeV via initial-state radiation (ISR)~\cite{conjugated}. In this process, a charmoniumlike state decaying to $D^+_sD_{s1}(2536)^-$ is observed for the first time. The data used in this analysis corresponds to $921.9~\infb$ of integrated luminosity at C.M. energies of 10.52, 10.58, and 10.867 GeV collected by the Belle detector~\cite{Belle1} at the KEKB asymmetric-energy $e^+e^-$ collider~\cite{KEKB1,KEKB2}.

We use {\sc phokhara}~\cite{PHOKHARA} to generate signal Monte Carlo (MC) events, determine the detector efficiency, and optimize selection criteria for signal events. Generic MC samples of $\yfos \to \bbc/\bbn$, $\yfis \to B^{(*)}_s\bar{B}^{(*)}_s$, and $\EE \to q\bar{q}~(q =u,~d,~s,~c)$ at $\sqrt{s} = 10.52,~10.58$, and $10.867~\gev$ with four times the luminosity of data are used to study possible backgrounds.

We fully reconstruct the ISR photon $\gamma_{\rm ISR}$, $D^+_s$, and $K^-$/$K^0_S$, but do not reconstruct the ${\bar D}^{*0}/D^{*-}$. Since the ${\bar D}^{*0}/D^{*-}$ decays are not reconstructed, the detector efficiency
for the $e^+e^-\to D^+_sD_{s1}(2536)^-(\to {\bar D}^{*0}K^-/D^{*-}K^0_S)$ process is greatly improved. For the measurement of the $e^+e^-\to D^+_sD_{s1}(2536)^-$ cross section,
we determine the invariant mass spectrum of $D^+_sD_{s1}(2536)^-$ ($M(D^+_sD_{s1}(2536)^-)$), which is equivalent to the mass recoiling against $\gamma_{\rm ISR}$ ($M_{\rm rec}(\gamma_{\rm ISR})$).
Here, $M_{\rm rec}(\gamma_{\rm ISR})$ is calculated using
$
M_{\rm rec}(\gamma_{\rm ISR}) = \sqrt{(P_{\rm C.M.}-P_{\gamma_{\rm ISR}})^2},
$
where $P_{\rm C.M.}$ and $P_{\gamma_{\rm ISR}}$ are the four-momenta of the initial $e^+e^-$ system and the ISR photon, respectively. However, the energy resolution of $\gamma_{\rm ISR}$ is very poor due to its high energy.
We constrain the recoil mass of the $\gamma_{\rm
ISR}D^+_sK^-$/$\gamma_{\rm ISR}D^+_sK^0_S$ to
the nominal mass of the ${\bar D}^{*0}$/$D^{*-}$ meson~\cite{PDG} to improve the resolution for the ISR photon in the events within the ${\bar D}^{*0}$/$D^{*-}$ signal region.
Before applying the mass constraint, the mass resolution of the $M(D^+_sD_{s1}(2536)^-)$ system is about 180~MeV/$c^2$. As a result of the constraint, the mass resolution is significantly improved, to about 5 MeV/c$^2$.

The $D^+_s$ candidates are reconstructed using eight decay modes: $\phi\pi^+$, ${\bar K}^*(892)^0K^+$, $K^0_SK^+$, $K^+K^-\pi^+\pi^0$, $K^0_S\pi^0K^+$, $K^*(892)^+K^0_S$, $\eta \pi^+$, and $\eta^{\prime}\pi^+$.
We use the techniques of Ref.~\cite{D092015} to reconstruct particles such as photons, charged pions and kaons, and $K_{S}^{0}$. The $\phi$, ${\bar K}^*(892)^0$, and $K^*(892)^+$ candidates are reconstructed in the $K^+K^-$, $K^-\pi^+$, and $K^0_S\pi^+$ decay modes. The invariant masses of the $K_{S}^{0}$, $\phi$, ${\bar K}^*(892)^0$, and $K^*(892)^+$ candidates are required to be within 10, 10, 50, and 50 MeV/$c^2$ of the corresponding nominal masses~\cite{PDG} ($>$95\% signal events are retained), respectively.

The most energetic ISR photon is required to have energy greater than 3 GeV in the $\EE$ C.M. frame.
Pairs of photons are combined to form $\pi^0$ candidates. The energies of the photons
from $\pi^0$ are required to be greater than 50 MeV in the calorimeter barrel
and 100 MeV in the calorimeter endcaps~\cite{ECL} in the laboratory frame. The $\eta$ candidates are reconstructed via
$\gamma\gamma$ and $\pi^+\pi^-\pi^0$ decay modes. Photon
candidates from $\eta\to\gamma\gamma$ are required to have
energies greater than 100~MeV in the laboratory frame. The
reconstructed $\eta$ candidates are then combined with $\pi^+\pi^-$ pairs to
form $\eta^{\prime}$ candidates.
The mass windows applied for $\pi^0$, $\eta\to\gamma\gamma$, $\eta\to\pi^+\pi^-\pi^0$, and
$\eta^{\prime}$ candidates are $\pm$12, $\pm$20, $\pm$10, and $\pm$10 MeV/$c^2$, which are within approximately 2.5$\sigma$ of the corresponding meson nominal masses~\cite{PDG}. After applying the mass window requirements, mass-constrained fits are applied to the $\pi^0$, $\eta$, and $\eta^{\prime}$ candidates to improve their momentum resolutions.

Before calculation of the $D_s^+$ candidate mass, a fit to a common vertex is performed for charged tracks in the $D_s^+$ candidate. After the application of the above requirements,  $D^+_s$ signals are clearly observed. We define the $D^+_s$ signal region as $|M(D^+_s)-m_{D^+_s}|<12$~MeV/$c^2$ ($\sim$2.0$\sigma$).
Here and throughout the text, $m_{i}$ represents the nominal mass of particle~$i$~\cite{PDG}.
To improve the momentum resolution of the $D^+_s$ meson candidate, a mass-constrained fit to the $D^+_s$ nominal mass~\cite{PDG} is performed. The $D^+_s$ mass sideband regions are defined as $1912.34<M(D^+_s)<1936.34$ MeV/c$^2$ and $2000.34<M(D^+_s)<2024.34$ MeV/c$^2$, which are twice as wide as the signal region.
The $D^+_s$ candidates from the sidebands are also constrained to the central mass values in the defined $D^+_s$ sideband regions. The $D^+_s$ candidate with the smallest $\chi^{2}$ from the $D^+_s$ mass fit is kept. Besides the selected ISR photon and $D^+_s$, we require at least one additional $K^-$ or $K^0_S$ candidate in the event, and retain all the combinations (the fraction of events with multiple candidates is 1.7\%).

\begin{figure}[htbp]
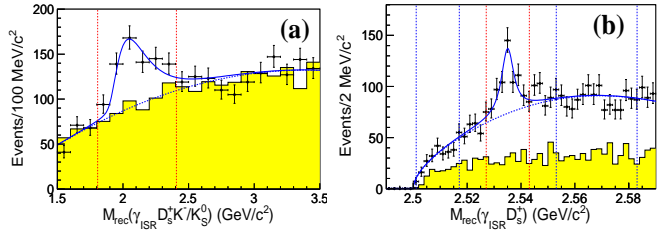

\vspace{0.3cm}
\includegraphics[height=4.25cm,width=3cm,angle=-90]{plot/fig1a.epsi}
\vspace{0.3cm}
\includegraphics[height=4.25cm,width=3cm,angle=-90]{plot/fig1b.epsi}
\caption{(a) The recoil mass spectrum against the $\gamma_{\rm ISR}D^+_sK^-/K^0_S$ system before applying the ${\bar D}^{*0}/D^{*-}$ mass constraint. The yellow histogram shows the normalized $D_{s1}(2536)^-$ mass sidebands (see below). The red dashed lines show the required ${\bar D}^{*0}/D^{*-}$ signal region. (b) The recoil mass spectrum
against the $\gamma_{\rm ISR}D^+_s$ system in
data. The yellow histogram shows the normalized $D^+_s$ mass
sidebands. The red dashed lines show the required $D_{s1}(2536)^-$ signal region,
and the blue dashed lines show the $D_{s1}(2536)^-$ mass sidebands.}\label{Ds1}
\end{figure}

Figure~\ref{Ds1}(a) shows the sum of the recoil mass spectra against the $\gamma_{\rm ISR}D^+_sK^-$ and $\gamma_{\rm ISR}D^+_sK^0_S$ systems after requiring the events within the $D_{s1}(2536)^-$ signal region (see below) in data.
Due to the poor recoil mass resolution, the ${\bar D}^{*0}/D^{*-}$ signal is very wide.
The ${\bar D}^{*0}/D^{*-}$ signal component is modeled using a Gaussian function
convolved with a Novosibirsk function~\cite{Nov} derived from the signal MC samples, while the combinatorial backgrounds are described by a second-order polynomial. The solid line is the total
fit and the ${\bar D}^{*0}/D^{*-}$ signal yield is $275\pm32$.
We define an asymmetric requirement of $-200$ $<$
$M_{\rm rec}(\gamma_{\rm
ISR}D^+_sK^-/K^0_S)-m_{{\bar D}^{*0}/D^{*-}}$ $<$ 400 MeV/$c^{2}$
for the ${\bar D}^{*0}/D^{*-}$ signal region.
Hereinafter the ${\bar D}^{*0}/D^{*-}$ mass constraint is applied
for events in the ${\bar D}^{*0}/D^{*-}$ signal region
to improve mass resolution.

The recoil mass spectrum against the $\gamma_{\rm ISR}D^+_s$ system after requiring the events within ${\bar D}^{*0}/D^{*-}$ signal region is shown in Fig.~\ref{Ds1}(b). A clear $D_{s1}(2536)^-$ signal is observed. The signal shape is described by a double Gaussian function (all the parameters are fixed to those from a fit to the MC simulated distribution), and a threshold function is used for the backgrounds. The threshold function is $(M_{\rm rec}-x_{\rm thr})^{\alpha}$$e^{[\beta_{1}(M_{\rm rec}-x_{\rm thr})+\beta_{2}(M_{\rm rec}-x_{\rm thr})^{2}]}$, where $M_{\rm rec}$ is the recoil mass of the $\gamma_{\rm ISR}D^+_s$; the parameters $\alpha$, $\beta_{1}$, and $\beta_{2}$ are free; the threshold parameter $x_{\rm thr}$ is fixed from generic MC simulations.
The fit yields $254\pm36$ $D_{s1}(2536)^-$ signal events as shown
in Fig.~\ref{Ds1}(b). We define the
$D_{s1}(2536)^-$ signal region as $|M_{\rm rec}(\gamma_{\rm
ISR}D^+_s)-m_{D_{s1}(2536)^-}|<8$~MeV/$c^2$ ($\sim 2.5\sigma$),
and sideband regions as shown by blue dashed
lines, which are three times as wide as the signal region. To
estimate the signal significance of the $D_{s1}(2536)^-$, we compute
$\sqrt{-2\ln(\mathcal{L}_0/\mathcal{L}_{\rm max})}$~\cite{significance},
where $\mathcal{L}_0$ and $\mathcal{L}_{\rm max}$ are the maximized likelihoods without and with the $D_{s1}(2536)^-$ signal, respectively. The statistical significance of the $D_{s1}(2536)^-$
signal is $8\sigma$.

\begin{figure}[htbp]
\vspace{0.2cm}
\includegraphics[height=7.5cm,width=8.5cm,angle=-90]{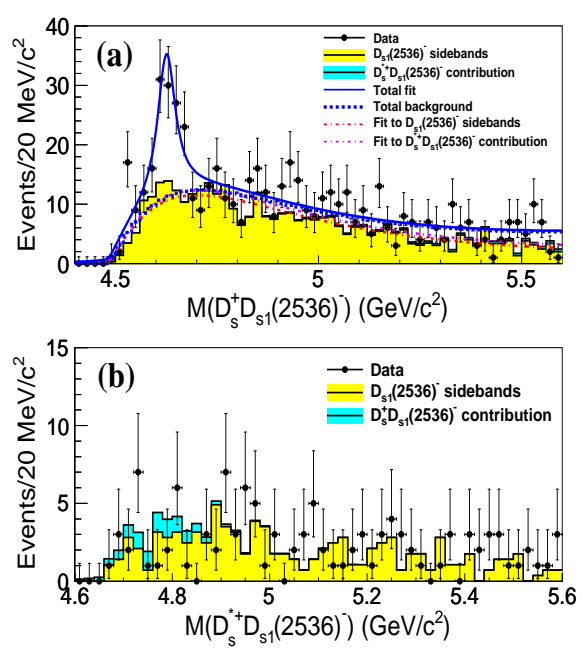}
\caption{(a) The $D^{+}_{s}D_{s1}(2536)^-$ invariant mass spectrum for $e^{+}e^{-}\to D^{+}_{s}D_{s1}(2536)^-$. (b) The $D^{*+}_{s}D_{s1}(2536)^-$ invariant mass spectrum for $e^{+}e^{-}\to D^{*+}_{s}D_{s1}(2536)^-$.
All the components including those from the fit to the $D^{+}_{s}D_{s1}(2536)^-$ invariant mass spectrum
are indicated in the labels and described in the text. Note that the cyan shaded histograms in the top/bottom show the $D^{+}_{s}D_{s1}(2536)^-$/$D^{*+}_{s}D_{s1}(2536)^-$ invariant mass spectrum from $D^{*+}_{s}D_{s1}(2536)^-$/$D^{+}_{s}D_{s1}(2536)^-$ background contribution after applying the requirements to reconstruct $e^{+}e^{-}\to D^{+}_{s}D_{s1}(2536)^-$/$e^{+}e^{-}\to D^{*+}_{s}D_{s1}(2536)^-$.
}\label{DssDs1}
\end{figure}

The $D^{+}_{s}D_{s1}(2536)^-$ invariant mass distribution is shown in Fig.~\ref{DssDs1}(a).
There is a significant peak around 4626 MeV/$c^2$, while no structure is seen in the normalized
$D_{s1}(2536)^-$ mass sidebands shown as the yellow histogram. In addition, no peaking background is found in the $D^+_sD_{s1}(2536)^-$ mass distribution from generic MC samples. We therefore interpret the peak in the data as evidence for an exotic charmoniumlike state decaying into $D^{+}_{s}D_{s1}(2536)^-$,  called $Y(4626)$ hereafter.
	
One possible background, which is not included in the $D_{s1}(2536)^-$ mass sidebands
is from $e^+e^-\to D^{*+}_{s}(\to D^+_s\gamma)D_{s1}(2536)^-$, where the photon from the $D^{*+}_{s}$ remains undetected.
To estimate such a background contribution, we measure this process with the data following the same procedure as used for the signal process. We require an extra photon with $E_{\gamma}>50$ MeV in the barrel or $E_{\gamma}>100$ MeV in the endcaps to combine with the $D^+_s$ to form the $D^{*+}_s$ candidate.
The mass and vertex fits are applied to the $D^{*+}_s$ candidates to improve their momentum resolution.
In events with multiple candidates, the best candidate is chosen using the lowest $\chi^2$ value from the mass-constrained fit. The same ${\bar D}^{*0}/D^{*-}$ signal region requirement on $M_{\rm rec}(\gamma_{\rm ISR}D^{*+}_sK^-/K^0_S)$ and the ${\bar D}^{*0}/D^{*-}$ mass constraint are applied as before in $e^+e^- \to D^+_sD_{s1}(2536)^-$. In the recoil mass spectrum of the $\gamma_{\rm ISR}D^{*+}_s$ an excess of events is observed in the $D_{s1}(2536)^-$ signal region.

After requiring the $D^+_sK^-/K^0_S$ mass to be within the $D_{s1}(2536)^-$ signal region, the $D^{*+}_{s}D_{s1}(2536)^-$ invariant mass distribution is shown in Fig.~\ref{DssDs1}(b). Note that the $e^{+}e^{-}\to D^{+}_{s}D_{s1}(2536)^-$ is a source of backgrounds for the $e^{+}e^{-}\to D^{*+}_{s}D_{s1}(2536)^-$ when the $D^{+}_{s}$ candidates are combined with soft photons to form $D^{*+}_{s}$ candidates. From Fig.~\ref{DssDs1}(b), no obvious structure is observed. The normalized contribution from $e^+e^-\to D^{*+}_{s}D_{s1}(2536)^-$ to $e^{+}e^{-}\to D^{+}_{s}D_{s1}(2536)^-$ is the cyan shaded histogram which is shown in Fig.~\ref{DssDs1}(a), and which is normalized to correspond to $N^{\rm obs}_{D^{*+}_{s}D_{s1}(2536)^-}\varepsilon_{D^{+}_{s}D_{s1}(2536)^-}/\varepsilon_{D^{*+}_{s}D_{s1}(2536)^-}$ events. Here, $N^{\rm obs}_{D^{*+}_{s}D_{s1}(2536)^-}$ is the yield of  $e^+e^-\to D^{*+}_{s}D_{s1}(2536)^-$ signal events in each $M(D^{*+}_{s}D_{s1}(2536)^-)$ bin in data after subtracting the normalized $D_{s1}(2536)^-$ sidebands and the $e^+e^-\to D^+_{s}D_{s1}(2536)^-$ background contribution, and $\varepsilon_{D^{+}_{s}D_{s1}(2536)^-}$ and $\varepsilon_{D^{*+}_{s}D_{s1}(2536)^-}$ are the reconstruction efficiencies for  $e^+e^-\to D^{+}_{s}D_{s1}(2536)^-$ and $e^+e^-\to D^{*+}_{s}D_{s1}(2536)^-$, respectively, and the ratio of efficiencies is $(1.00\pm0.02)$. The yield of $D^{*+}_{s}D_{s1}(2536)^-$ after the background subtraction for the whole region in Fig.~\ref{DssDs1}(b) is $(11.6\pm3.6)$. A similar method is applied to estimate the background contribution from $e^{+}e^{-}\to D^+_{s}D_{s1}(2536)^-$ to $e^{+}e^{-}\to D^{*+}_{s}D_{s1}(2536)^-$.

We perform an unbinned likelihood fit simultaneously to the $M(D^+_sD_{s1}(2536)^-)$
distributions of all selected $D_{s1}(2536)^-$ signal candidates, the normalized $D_{s1}(2536)^-$ mass
sidebands, and the $e^+e^-\to D^{*+}_{s}D_{s1}(2536)^-$ contribution.
The following components are included in the fit to the $M(D^+_sD_{s1}(2536)^-)$ distribution: a resonance signal, a non-resonant contribution, the $D_{s1}(2536)^-$ mass sidebands, and the $e^+e^-\to D^{*+}_{s}D_{s1}(2536)^-$ contribution. A Breit-Wigner (BW) function convolved with a Gaussian function (with its width fixed at 5.0~MeV/c$^{2}$ according to the MC simulation), multiplied by an efficiency function that has a linear dependence on $M(D^{+}_{s}D_{s1}(2536)^-)$ and the differential ISR effective luminosity~\cite{lum} is taken as the signal shape. Here the BW formula used has the form~\cite{D95.092007}
\begin{equation} \label{eq:q1}
BW(\sqrt{s})=\frac{\sqrt{12\pi\Gamma_{ee}\BR_f\Gamma}}{s-M^2+iM\Gamma}\sqrt{\frac{\Phi_2(\sqrt{s})}{\Phi_2(M)}},
\end{equation}
where $M$ is the mass of the resonance, $\Gamma$ and $\Gamma_{ee}$ are the total width and partial width to $e^+e^-$, $\BR_f$ = $\BR(Y(4626) \to D^+_sD_{s1}(2536)^-)\times\BR(D_{s1}(2536)^-\to {\bar D}^{*0}K^{-})$ is the product branching fraction of the $Y(4626)$ into the final state, and $\Phi_2$ is the two-body decay phase space factor that increases smoothly from the mass threshold with $\sqrt{s}$. A two-body phase space form is also taken into account for the non-resonant contribution.
The $D_{s1}(2536)^-$ mass sidebands and the $e^+e^-\to D^{*+}_{s}D_{s1}(2536)^-$ contribution are parameterized with threshold functions.

The fit results are shown in Fig.~\ref{DssDs1}(a), where the solid blue curve is the best fit,
the blue dotted curve is the sum of the backgrounds, the red dot-dashed curve is the fitted result to the normalized $D_{s1}(2536)^-$ mass sidebands, and the violet dot-dashed curve is for the $e^+e^-\to D^{*+}_{s}D_{s1}(2536)^-$ contribution. The yield of the $Y(4626)$ signal is $89^{+17}_{-16}$. The statistical significance of the $Y(4626)$ signal is $6.5\sigma$, calculated from the difference of the logarithmic likelihoods~\cite{significance}, $-2\ln(\mathcal{L}_{0}/\mathcal{L}_{\rm max}) = 50.4$, where $\mathcal{L}_{0}$ and $\mathcal{L}_{\rm max}$ are the maximized likelihoods without and with a signal component, respectively, taking into account the difference in the number of degrees of freedom ($\Delta$ndf = 3). The parameterization of the non-resonant contribution is the dominant systematic uncertainty for the estimate of the signal significance. Changing the two-body phase space form to a threshold function parameterized by $\sqrt{M-x_{\rm thr}}$ or a two-body phase space form plus a threshold function, the $Y(4626)$ signal significance is reduced to 5.9$\sigma$. We take this value as the signal significance with systematic uncertainties included. The fitted mass and width for the $Y(4626)$ are $(4625.9^{+6.2}_{-6.0}({\rm stat.})\pm0.4({\rm syst.}))~{\rm MeV}/c^{2}$ and $(49.8^{+13.9}_{-11.5}({\rm stat.})\pm4.0({\rm syst.}))~{\rm MeV}$, respectively. The value of $\Gamma_{ee}\times\BR(Y(4626) \to D^+_sD_{s1}(2536)^-)\times\BR(D_{s1}(2536)^-\to {\bar D}^{*0}K^{-})$ is obtained to be $(14.3^{+2.8}_{-2.6}(\rm stat.)\pm1.5(syst.))$ eV. The systematic uncertainties are discussed below.

\begin{figure}[htbp]
\vspace{0.2cm}
\includegraphics[height=6.5cm,angle=-90]{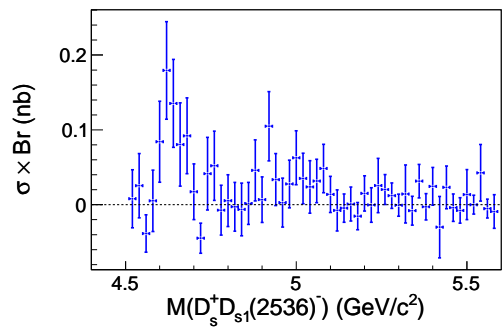}
\caption{The product of the $e^+e^-\to D^+_sD_{s1}(2536)^-$ cross section and branching fraction
$\BR(D_{s1}(2536)^-\to {\bar D}^{*0}K^-)$ as a function of $M(D^+_sD_{s1}(2536)^-)$; here the statistical and systematic uncertainties are summed in quadrature.}\label{CS}
\end{figure}

The product of the $e^+e^-\to D^+_sD_{s1}(2536)^-$ dressed cross section ($\sigma$)~\cite{dressed} and the decay branching fraction $\BR(D_{s1}(2536)^-\to {\bar D}^{*0}K^-)$ for each $D^+_sD_{s1}(2536)^-$ mass bin
from threshold to 5.59 GeV/$c^2$ in steps of 20 MeV/$c^2$ is computed as
\begin{equation} \label{eq:q2}
\begin{aligned}
&\sigma(e^+e^-\to D^+_sD_{s1}(2536)^-)\BR(D_{s1}(2536)^-\to {\bar D}^{*0}K^-) = \\
&\frac{N^{D_{s1}(2536)^-}_{\rm fit}}{ {d\cal{L}} \times  [\Sigma_{i}(\varepsilon^{{\bar D}^{*0}K^-}_{i} \times \BR_{i}) + R^{D^{*-}K^0_S}_{{\bar D}^{*0}K^-}\times\Sigma_{i}(\varepsilon^{D^{*-}K^0_S}_{i} \times \BR_{i})]}, \\
\end{aligned}
\end{equation}
where $N^{D_{s1}(2536)^-}_{\rm fit}$, $d{\cal{L}}$, $\Sigma_{i}(\varepsilon^{{\bar D}^{*0}K^-}_{i} \times \BR_{i})$, and $\Sigma_{i}(\varepsilon^{D^{*-}K^0_S}_{i} \times \BR_{i})$
are the yields of fitted $D_{s1}(2536)^-$ signal events after subtracting the $e^+e^-\to D^{*+}_{s}D_{s1}(2536)^-$ background contribution in data, the effective luminosity~\cite{lum}, the sums of the product of the reconstruction efficiency and branching
fraction for each $D^+_s$ decay mode ($i$) in $D_{s1}(2536)^-\to {\bar D}^{*0}K^-$ and $D_{s1}(2536)^-\to D^{*-}K^0_S$,
in each $D^{+}_{s}D_{s1}(2536)^-$ mass bin, respectively;
$R^{D^{*-}K^0_S}_{{\bar D}^{*0}K^-} = \BR(D_{s1}(2536)^-\to D^{*-}K^0_S)/\BR(D_{s1}(2536)^-\to {\bar D}^{*0}K^-)=0.425\pm0.06$
is taken from Ref.~\cite{PDG}. The values used to calculate $\sigma(e^+e^-\to D^+_sD_{s1}(2536)^-)\BR(D_{s1}(2536)^-\to {\bar D}^{*0}K^-)$ are summarized in the Supplemental Material~\cite{SM}. In the fit to the recoil mass spectrum of $\gamma_{\rm ISR}D^{+}_{s}$ combinations in each $D^{+}_{s}D_{s1}(2536)^-$ mass bin, the $D_{s1}(2536)^-$ signal shape is fixed to that from the overall fit, as shown
by the blue solid curve in Fig.~\ref{Ds1}, and a threshold function is used for the backgrounds.
The resulting $\sigma(e^{+}e^{-}\to D^{+}_{s}D_{s1}(2536)^-) \times \BR(D_{s1}(2536)^-\to {\bar D}^{*0}K^{-})$ value as a function of $M(D^+_sD_{s1}(2536)^-)$ is shown in Fig.~\ref{CS} with the statistical and systematic uncertainties (discussed below) summed in quadrature.

The sources of systematic uncertainties for the cross section measurement
include detection-efficiency-related uncertainties, branching fractions of the intermediate states,
fit uncertainty, resonance parameters, the MC event generator, $e^+e^-\to D^{*+}_{s}D_{s1}(2536)^-$ background contribution as well as the integrated luminosity.
The detection-efficiency-related uncertainties include those for tracking efficiency
(0.35\%/track), particle identification efficiency (1.1\%/kaon and 0.9\%/pion), $\ks$ selection efficiency
(1.4\%)~\cite{L119}, $\pi^0$ reconstruction efficiency (2.25\%/$\pi^0$), and photon reconstruction
efficiency (2.0\%/photon).
The above individual uncertainties from different $D^+_s$ decay channels are added linearly, weighted by the product of the detection efficiency and $D^+_s$ partial decay width. These uncertainties are summed in quadrature to obtain the final uncertainty related to the reconstruction efficiency.
Uncertainties for $D^{+}_s$ decay branching fractions and $R^{D^{*-}K^0_S}_{{\bar D}^{*0}K^-}$
are taken from Ref.~\cite{PDG}; the final uncertainties on the $D^{+}_s$ partial decay widths
are summed in quadrature over the eight $D^{+}_s$ decay modes weighted by the product of the
efficiency and the $D^+_s$ partial decay width.
Systematic uncertainties associated with the fitting procedure are estimated by
changing the order of the background polynomial and the range of the fit.
The deviations from nominal fit results are taken
as systematic uncertainties. Changing the values of mass and width of $D_{s1}(2536)^-$ by 1$\sigma$~\cite{PDG} in each $M(D^+_sD_{s1}(2536)^-)$ bin has no effect on the fits. Thus, the uncertainty from the resonance parameters can be neglected.
The {\sc phokhara} generator calculates the ISR-photon radiator function with 0.1\% accuracy~\cite{PHOKHARA}. The uncertainty attributed to the generator can also be neglected.

By fitting the $D_{s1}(2536)^-$ mass spectrum in each $M(D^{*+}_sD_{s1}(2536)^-)$ bin for $e^+e^-\to D^{*+}_sD_{s1}(2536)^-$, we find the signal yields are less than 1. In addition, the $D_{s1}(2536)^-$ signal from the $e^+e^-\to D^{*+}_{s}D_{s1}(2536)^-$ contribution has a much poorer mass resolution according to MC simulation. Therefore, the systematic uncertainty associated with the $e^+e^-\to D^{*+}_{s}D_{s1}(2536)^-$ contribution is neglected.
The total luminosity is determined to 1.4\% precision using wide-angle Bhabha scattering events.
All the uncertainties are summarized in Table~\ref{systematic}. Assuming all the sources are independent,
we sum them in quadrature to obtain the total systematic uncertainties.

\linespread{1.2}
\begin{table}[htbp]
\caption{Summary of the absolute systematic uncertainties ($\sigma_{\rm sys}$)
on the product of $e^+e^-\to D^+_sD_{s1}(2536)^-$ cross section and the decay branching fraction
$\BR(D_{s1}(2536)^-\to {\bar D}^{*0}K^-)$ for different $M(D^+_sD_{s1}(2536)^-)$ bins.}
\vspace{0.2cm}
\label{systematic}
\begin{tabular}{c || c }
\hline
~~Source~~ & ~~~$\sigma_{\rm sys}$ (pb)~~~ \\\hline
Detection efficiency & $0.0-8.1$ \\
Branching fractions & $0.0-7.4$ \\
Fit uncertainty &  $0.5-36.2$ \\
Luminosity & $0.0-2.5$ \\\hline
Quadratic Sum &  $0.6-36.3$ \\\hline
\end{tabular}
\end{table}

The following systematic uncertainties on the measured mass and width for the $Y(4626)$, and the $\Gamma_{ee}\times\BR(Y(4626) \to D^+_sD_{s1}(2536)^-)\times\BR(D_{s1}(2536)^-\to {\bar D}^{*0}K^{-})$ are considered.
MC simulation is known to reproduce the resolution of mass peaks within 10\% over a large number of different systems. The resultant systematic uncertainty in the width and $\Gamma_{ee}\times\BR(Y(4626) \to D^+_sD_{s1}(2536)^-)\times\BR(D_{s1}(2536)^-\to {\bar D}^{*0}K^{-})$ from this source is 0.3 MeV and 0.1 eV. By changing the non-resonant background shape to a threshold function or to the sum of a two-body phase space form and a threshold function, the differences of 0.3 MeV/$c^2$ and 3.9 MeV in the measured mass and width, and 1.3 eV for the $\Gamma_{ee}\times\BR(Y(4626) \to D^+_sD_{s1}(2536)^-)\times\BR(D_{s1}(2536)^-\to {\bar D}^{*0}K^{-})$, respectively, are taken as systematic uncertainties.
The uncertainty in the efficiency correction from detection efficiency, branching fractions of the intermediate states, and the integrated luminosity is 4.9\%. Changing the efficiency function by 4.9\% gives a 0.1 MeV/$c^2$ change on the mass, 0.2 MeV on the width, and 0.7 eV on the product $\Gamma_{ee}\times\BR(Y(4626) \to D^+_sD_{s1}(2536)^-)\times\BR(D_{s1}(2536)^-\to {\bar D}^{*0}K^{-})$. Finally, the total systematic uncertainties on the $Y(4626)$ mass, width, and $\Gamma_{ee}\times\BR(Y(4626) \to D^+_sD_{s1}(2536)^-)\times\BR(D_{s1}(2536)^-\to {\bar D}^{*0}K^{-})$ are 0.4 MeV/$c^2$, 4.0 MeV, and 1.5 eV, respectively.

In summary, the product of the $e^+e^-\to D^+_sD_{s1}(2536)^-$ cross section and the decay branching fraction
$\BR(D_{s1}(2536)^-\to {\bar D}^{*0}K^-)$ is measured over the C.M. energy range from the $D^+_sD_{s1}(2536)^-$ mass threshold to 5.59~GeV for the first time. We observe the first vector charmoniumlike state decaying to a charmed-antistrange and anticharmed-strange meson pair $D^+_sD_{s1}(2536)^-$ with a signal significance of 5.9$\sigma$ with systematic uncertainties included. The measured mass and width are $(4625.9^{+6.2}_{-6.0}({\rm stat.})\pm0.4({\rm syst.}))~{\rm MeV}/c^{2}$ and $(49.8^{+13.9}_{-11.5}({\rm stat.})\pm4.0({\rm syst.}))~{\rm MeV}$, respectively, which are close to those of the
$Y(4660)$ state~\cite{PDG}. The $\Gamma_{ee}\times\BR(Y(4626) \to D^+_sD_{s1}(2536)^-)\times\BR(D_{s1}(2536)^-\to {\bar D}^{*0}K^{-})$ is obtained to be $(14.3^{+2.8}_{-2.6}(\rm stat.)\pm1.5(syst.))$ eV.


We thank the KEKB group for excellent operation of the
accelerator; the KEK cryogenics group for efficient solenoid
operations; and the KEK computer group, the NII, and
PNNL/EMSL for valuable computing and SINET5 network support.
We acknowledge support from MEXT, JSPS and Nagoya's TLPRC (Japan);
ARC (Australia); FWF (Austria); NSFC and CCEPP (China);
MSMT (Czechia); CZF, DFG, EXC153, and VS (Germany);
DST (India); INFN (Italy);
MOE, MSIP, NRF, RSRI, FLRFAS project, GSDC of KISTI and KREONET/GLORIAD (Korea);
MNiSW and NCN (Poland); MSHE, Agreement 14.W03.31.0026 (Russia); ARRS (Slovenia);
IKERBASQUE (Spain); SNSF (Switzerland); MOE and MOST (Taiwan); and DOE and NSF (USA).

\renewcommand{\baselinestretch}{1.2}

\newpage

\begin{minipage}{2\linewidth}
\Large	
\centering
{\bf Supplemental Material}
\vspace{1.0cm}

\large
\centering
{\bf  The product of cross section and decay branching fraction $\sigma(e^+e^-\to D^+_sD_{s1}(2536)^-)\times\BR(D_{s1}(2536)^-\to {\bar D}^{*0}K^-)$}
\vspace{0.5cm}

\raggedright
\normalsize
~~~The number of effective luminosity $d\cal L$, the total reconstruction efficiency $\varepsilon_{\rm tot}$, the number of fitted signal events $N^{\rm fit}$, and the product of dressed cross section and decay branching fraction $\sigma\times\BR$ in each $D^+_sD_{s1}(2536)^-$ mass bin are summarized in Table~\ref{table1}. The $\varepsilon_{\rm tot}$ equals to $[\Sigma_{i}(\varepsilon^{{\bar D}^{*0}K^-}_{i} \times \BR_{i}) + R^{D^{*-}K^0_S}_{{\bar D}^{*0}K^-}\times\Sigma_{i}(\varepsilon^{D^{*-}K^0_S}_{i} \times \BR_{i})]$, where $\Sigma_{i}(\varepsilon^{{\bar D}^{*0}K^-}_{i} \times \BR_{i})$ and $\Sigma_{i}(\varepsilon^{D^{*-}K^0_S}_{i} \times \BR_{i})$ are the sums of the product of reconstruction efficiency and branching fraction for each $D^+_s$ decay mode ($i$) in $D_{s1}(2536)^-\to {\bar D}^{*0}K^-$ and $D_{s1}(2536)^-\to D^{*-}K^0_S$, and $R^{D^{*-}K^0_S}_{{\bar D}^{*0}K^-} = \BR(D_{s1}(2536)^-\to D^{*-}K^0_S)/\BR(D_{s1}(2536)^-\to {\bar D}^{*0}K^-)$ = $(0.425\pm0.06)$.

\end{minipage}
\linespread{1.1}
\begin{table*}[htbp]
\caption{The values of the product of dressed cross section and decay branching fraction $\sigma(e^+e^-\to D^+_sD_{s1}(2536)^-)\times\BR(D_{s1}(2536)^-\to {\bar D}^{*0}K^-)$ for the $e^+e^- \to D^+_sD_{s1}(2536)^-+c.c.$ varying from $\sqrt{s}$ = 4.51 to 5.59 GeV in a step of 0.02 GeV. For the $N^{\rm fit}$, the uncertainty is statistical only, while for the dressed Born cross section $\sigma$, the first uncertainty is statistical and the second is systematic.}
\vspace{0.2cm}
\label{table1}
\begin{tabular}{c | c | r@{$\pm$}l | c | r@{$\pm$} r@{$\pm$}l}
\hline\hline
\scriptsize ~~~$\sqrt{s}$ (GeV)~~~ & ~~~$d\cal L$ (pb$^{-1}$)~~~ & \multicolumn{2}{  c | }{$N^{\rm fit}$} & ~~~$\varepsilon_{\rm tot}$ (\%)~~~ & \multicolumn{3}{  c  }{$\sigma\times\BR$ (pb)}  \\\hline
4.52	&	87.6	&	0.9		&	4.3	&	0.128	&	8.0	&	38.8	&	0.7		\\
4.54	&	88.2	&	2.9		&	4.9	&	0.129	&	25.4	&	42.8	&	2.7		\\
4.56	&	88.8	&	$-$4.4		&	2.8	&	0.130	&	$-$38.5	&	24.3	&	5.7		\\
4.58	&	89.4	&	0.6		&	4.8	&	0.130	&	5.2	&	40.9	&	3.2		\\
4.60	&	90.0	&	9.9		&	6.0	&	0.131	&	84.2	&	50.9	&	18.4		\\
4.62	&	90.6	&	~~~21.4		&	7.3~~~	&	0.132	&	~~~179.4	&	61.5	&	21.4~~~		\\
4.64	&	91.3	&	16.4		&	6.5	&	0.132	&	135.3	&	54.0	&	23.1		\\
4.66	&	91.9	&	9.8		&	6.1	&	0.133	&	80.4	&	50.0	&	24.7		\\
4.68	&	92.5	&	11.4		&	5.4	&	0.134	&	92.1	&	43.5	&	25.8		\\
4.70	&	93.1	&	2.2		&	4.5	&	0.134	&	17.4	&	35.6	&	11.0		\\
4.72	&	93.8	&	$-$5.7		&	2.5	&	0.135	&	$-$44.7	&	19.6	&	4.1		\\
4.74	&	94.4	&	5.3		&	5.7	&	0.135	&	41.5	&	44.5	&	19.4		\\
4.76	&	95.0	&	6.7		&	5.8	&	0.136	&	52.1	&	44.9	&	11.2		\\
4.78	&	95.7	&	$-$1.0		&	4.3	&	0.136	&	$-$7.3	&	33.3	&	1.1		\\
4.80	&	96.3	&	0.7		&	4.5	&	0.137	&	5.4	&	33.7	&	7.7		\\
4.82	&	97.0	&	$-$0.4		&	3.5	&	0.137	&	$-$2.7	&	26.4	&	19.0		\\
4.84	&	97.7	&	$-$0.8		&	4.7	&	0.138	&	$-$6.3	&	34.6	&	6.9		\\
4.86	&	98.3	&	0.2		&	3.6	&	0.138	&	1.7	&	26.6	&	9.2		\\
4.88	&	99.0	&	6.3		&	5.6	&	0.139	&	45.9	&	40.7	&	3.6		\\
4.90	&	99.7	&	0.9		&	3.5	&	0.139	&	6.8	&	25.3	&	16.9		\\
4.92	&	100.4	&	14.7		&	6.3	&	0.140	&	104.9	&	45.0	&	10.8		\\
4.94	&	101.1	&	4.8		&	4.8	&	0.140	&	33.6	&	33.6	&	8.8		\\
4.96	&	101.7	&	0.4		&	4.5	&	0.141	&	2.8	&	31.1	&	10.1		\\
4.98	&	102.4	&	4.0		&	4.6	&	0.141	&	27.7	&	31.5	&	4.5		\\
5.00	&	103.1	&	9.1		&	5.0	&	0.141	&	62.7	&	34.1	&	11.5		\\
5.02	&	103.9	&	5.2		&	4.6	&	0.142	&	35.1	&	30.9	&	13.7		\\
5.04	&	104.6	&	3.5		&	4.3	&	0.142	&	23.8	&	29.3	&	19.2		\\
5.06	&	105.3	&	4.8		&	4.3	&	0.142	&	31.7	&	29.0	&	2.2		\\
5.08	&	106.0	&	7.3		&	4.8	&	0.143	&	48.4	&	31.4	&	8.0		\\
5.10	&	106.7	&	2.1		&	3.2	&	0.143	&	13.8	&	21.0	&	12.0		\\
5.12	&	107.5	&	$-$1.2		&	4.1	&	0.144	&	$-$7.6	&	26.8	&	6.1		\\
5.14	&	108.2	&	$-$0.7		&	2.8	&	0.144	&	$-$4.5	&	17.9	&	6.2		\\
5.16	&	109.0	&	0.1		&	3.5	&	0.144	&	0.9	&	22.4	&	4.1		\\
5.18	&	109.7	&	$-$2.4		&	2.5	&	0.145	&	$-$15.3	&	15.6	&	9.2		\\
5.20	&	110.5	&	2.4		&	3.7	&	0.145	&	15.1	&	23.2	&	4.4		\\
5.22	&	111.3	&	$-$0.1		&	3.4	&	0.146	&	$-$0.4	&	21.1	&	9.2		\\
5.24	&	112.0	&	4.2		&	4.2	&	0.146	&	25.6	&	25.9	&	25.4		\\
5.26	&	112.8	&	3.4		&	3.3	&	0.146	&	20.3	&	20.0	&	1.5		\\
5.28	&	113.6	&	2.0		&	2.8	&	0.147	&	12.2	&	16.7	&	2.8		\\
5.30	&	114.4	&	$-$0.1		&	2.5	&	0.147	&	$-$0.7	&	14.7	&	0.6		\\
5.32	&	115.2	&	2.4		&	4.0	&	0.148	&	14.3	&	23.7	&	30.6		\\
5.34	&	116.0	&	$-$1.4		&	3.3	&	0.148	&	$-$8.0	&	19.2	&	1.5		\\
5.36	&	116.8	&	5.5		&	3.8	&	0.149	&	31.5	&	22.0	&	3.7		\\
5.38	&	117.6	&	$-$0.5		&	3.0	&	0.149	&	$-$2.8	&	17.1	&	4.4		\\
5.40	&	118.5	&	4.4		&	3.7	&	0.150	&	24.5	&	21.0	&	13.9		\\
5.42	&	119.3	&	$-$5.4		&	3.4	&	0.150	&	$-$30.0	&	19.0	&	36.3		\\
5.44	&	120.1	&	4.2		&	4.1	&	0.151	&	23.2	&	22.8	&	17.1		\\
5.46	&	121.0	&	$-$0.7		&	3.0	&	0.151	&	$-$3.9	&	16.4	&	8.3		\\
5.48	&	121.9	&	$-$1.4		&	2.7	&	0.152	&	$-$7.7	&	14.5	&	8.9		\\
5.50	&	122.7	&	2.5		&	3.2	&	0.153	&	13.6	&	17.1	&	28.7		\\
5.52	&	123.6	&	0.0		&	2.4	&	0.153	&	0.0	&	12.6	&	0.8		\\
5.54	&	124.5	&	8.2		&	5.2	&	0.154	&	42.6	&	27.0	&	26.6		\\
5.56	&	125.4	&	$-$1.0		&	2.1	&	0.155	&	$-$5.2	&	10.7	&	6.7		\\
5.58	&	126.3	&	$-$1.8		&	1.8	&	0.155	&	$-$9.1	&	9.0	&	20.6		\\\hline\hline
\end{tabular}
\end{table*}


\begin{thebibliography}{**}
\bibitem{conjugated} Throughout this analysis, for any given mode, the corresponding charge-conjugate mode is implied.
\bibitem{Nov} The Novosibirsk function is defined as $f(x) =
\exp[-\frac12(\ln^2(1+\Lambda(x-x_0))/\tau^2+\tau^2)]$ with
$\Lambda=$ $\sinh(\tau\sqrt{\ln 4})/(\sigma\sqrt{\ln4})$. The
parameters represent the mean ($x_0$), the width ($\sigma$) and
the tail asymmetry ($\tau$).
\bibitem{dressed} The dressed cross section is $\sigma=\sigma^{B}/|1-\Pi|^2$, where $\sigma^{B}$ is the Born cross section, and $|1-\Pi|^2=0.94$ is the vacuum polarization factor~\cite{VP}.
\end{thebibliography}
\end{document}